\documentclass[reprint,superscriptaddress,nobibnotes,amsmath,amssymb,aps,prl,floatfix]{revtex4-2}
\usepackage{multirow}
\usepackage{graphicx}
\usepackage{hyperref}
\hypersetup{colorlinks=true, linkcolor=blue, filecolor=blue, urlcolor=blue, citecolor=blue }
\usepackage{dcolumn}
\usepackage{bm}
\usepackage[dvipsnames]{xcolor}
\usepackage{textcomp}
\newcommand{\etal}[1]{\textcolor{black}{{\textit{et al.}}}}

\begin{document}
\title{Ionic Self-Diffusion and the Glass Transition Anomaly in Aluminosilicates}
\author{Achraf Atila}
\email{achraf.atila@fau.de}
\affiliation{Friedrich-Alexander-Universit\"at Erlangen-N\"urnberg, Materials Science and Engineering, Institute I, Martensstr. 5, Erlangen 91058, Germany}
\author{Said Ouaskit}
\email{said.ouaskit@univh2c.ma}
\affiliation{Laboratoire physique de la mati\'ere condens\'ee, Facult\'e des sciences Ben M\textquotesingle sik, Universit\'e Hassan II de Casablanca, B.P 7955, Av Driss El Harti, Sidi Othmane, Casablanca, Maroc}
\author{Abdellatif Hasnaoui}
\email{abdellatif.hasnaoui@uhp.ac.ma}
\affiliation{LS3M, Facult\'e Polydisciplinaire Khouribga, Sultan Moulay Slimane University of Beni Mellal, B.P 145, 25000 Khouribga, Morocco}
\date{\today}

\begin{abstract}
The glass transition temperature is the temperature, after which the supercooled liquid undergoes a dynamical arrest. Usually, the glass network modifiers (e.g., Na$_2$O) affect the behavior of T$_g$. However, in aluminosilicate glasses, the effect of different modifiers on T$_g$ is still unclear and show an anomalous behavior. Here, based on molecular dynamics simulations, we show that the glass transition temperature (T$_g$) decreases with increasing charge balancing cations field strength (FS) in the aluminosilicate glasses, which is an anomalous behavior as compared to other oxide glasses. The results show that the origins of this anomaly come from the dynamics of the supercooled liquid above T$_g$, which in turn is correlated to pair excess entropy. Our results deepen our understanding of the effect of different modifiers on the properties of the aluminosilicate glasses.
\end{abstract}
\maketitle
\textbf{\textit{Introduction.---}}The aluminosilicate glasses are of great interest in glass science and industry. This is due to their good physical properties and chemical stability, without neglecting the abundance of the glass elements in the earth's crust \cite{Yu2015, Welch2013, Atila2019b}. The glass transition temperature is a vital glass property when considering glasses for a specific application. T$_g$ is the temperature; below it, the physical properties of the supercooled liquid change to those of a glassy state presenting rigidity and elasticity due to the structural dynamical arrest. Knowing which factors control T$_g$ and how different modifiers affect the values of T$_g$ will deepen our physical and chemical understanding of the behavior of the glass transition.

Cation field strength (FS) as defined by Dietzel \cite{Dietzel1942} FS = Z$_{X}$/(r$_{X}+$r$_{O}$)$^2$, where Z and r stand for the charge and ionic radius, respectively, and gives an indication of the cation-oxygen bond strength (higher FS means higher bond strength). This parameter could be used to study the effect of different modifiers or charge balancing cations on the properties of oxide glasses. Based on this assumption, it is expected that the properties of the studied oxide glasses to scale with FS.\\
The properties of oxide glasses such as T$_g$, hardness, and elastic moduli of binary alkali tellurite \cite{Wilkinson2018, Barney2015, Tagiara2019, Zhang1992}, phosphates \cite{Hermansen2014}, and ternary alkali/ alkaline earth aluminoborate glasses \cite{Januchta2017, Mascaraque2018} indeed scales with FS. These results suggested that the physical properties of the oxide glasses depend strongly on the charge balancing cations field strength. Furthermore, it has been found that the elastic properties increase with increasing cations field strength in binary alkali silicate glasses \cite{Pedone2008b}. Recently, we have highlighted the same effect in ternary alkali and alkaline earth aluminosilicate glasses \cite{Atila2019b}. Moreover, we showed that the behavior of T$_g$ does not correlate positively with FS and thus showing an anomalous behavior as compared to the previously mentioned glasses. This anomalous behavior of T$_g$ observed from our previous MD simulations \cite{Atila2019b} was already found experimentally by Weigel \textit{et al.} \cite{Weigel2016} and theoretically by Pedone \textit{et al.} in binary silicate glasses \cite{Pedone2008b}. The anomalous behavior of T$_g$ is still not explained in the literature and remains unclear. To harness this effect, we simulated eight charge-balanced aluminosilicate glasses using molecular dynamics (MD) simulations, which offers a powerful tool for investigating and understanding the atomic-scale behavior of materials.

In this letter, we investigate the origins of the anomalous behavior of T$_g$ in the charge balanced aluminosilicates. We show that this behavior is correlated to the diffusion behavior of the charge balancing cations and to the degree of the ordering in the glasses. The glasses contain either monovalent alkali cation, divalent alkaline earth cations, or Zn. The content of the modifier was set to be equal to that of alumina giving a ratio R=1 (R=[X$_{2/n}^{n+}$O]/[Al$_2$O$_3$]) \cite{Atila2019a, Atila2019b}. Based on these assumptions, the modifiers are expected to behave similarly in the aluminosilicate glasses network (to charge balance tetrahedral AlO$^4$ units), although, the charge balancing ability will be different ( due to the difference in the size of the cations) as the same charge is distributed over a larger area for larger cations. Moreover, the oxygen ionic radius is omitted from the FS equation as it is approximated to be the same (FS = $\frac{Z_X}{r_X^2}$).

\textbf{\textit{Methods.---}}In this work eight charge-balanced aluminosilicate glasses (X$_{n/2}^{n+}$O)$_{25}$-(Al$_{2}$O$_{3}$)$_{25}$-(SiO$_{2}$)$_{50}$ (X stands for Li, Na, K, Mg, Ca, Sr, Ba, or Zn) were simulated using classical molecular dynamics. The well-established potential by Pedone \textit{et al.} \cite{Pedone2006} was used to model the interactions between atoms. This potential gives a realistic agreement with available experimental data, as mentioned in the literature \cite{Yu2018, Atila2019b, Ghardi2019, Pedone2008b}. Potential parameters and partial charges are given in the reference \cite{Pedone2006}. All simulations were performed using the large-scale atomic/molecular massively parallel simulator LAMMPS \cite{Plimpton1995}. Velocity-Verlet algorithm with an integration time step of 1 fs was used to integrate the equations of motion. Periodic boundary conditions (PBC) are applied in all directions to avoid edge effects and to simulate bulk systems. Long-range interactions were evaluated by the Ewald summation method, with a real-space cutoff of 12.0 \AA\ and precision of $10^{-6}$. The short-range interactions cutoff distance was chosen to be 5.5 \AA\ \cite{Pedone2006}. 

All systems consist of approximately 4200 atoms placed randomly in a cubic simulation box, ensuring that there is no unrealistic overlap between atoms. First, we equilibrated the systems at a high temperature (T = 5000 K) for 500 ps. This step is needed to ensure that each system loses the memory of its initial configuration. After that, the systems were subsequently quenched linearly from the liquid temperature (T = 5000 K) to room temperature (T = 300 K) with a cooling rate of $10^{12}$ K/ps while keeping the volume fixed. Nos\'e-Hoover thermostat and barostat were used to control the temperature and pressure. At 300 K, another run for 1 ns and zero pressure in NPT ensemble was performed, and the structural (short- and medium-range structures) and elastic properties were found to be in a realistic agreement with the experimental data as discussed in the reference \cite{Atila2019b}. Cooling rates used in molecular dynamics are much higher than those used in experiments due to the intrinsic incapability of molecular dynamics to use very low cooling rates. The values of the cooling rate used in the present simulations are usually used in making glasses in MD simulations, and changing these values over an order of magnitude does not affect the physical properties and the short-range structure of the glassy state considerably \cite{Atila2019b, li2017cooling}.

\begin{figure}[ht!]
\centering
\includegraphics[width=\columnwidth]{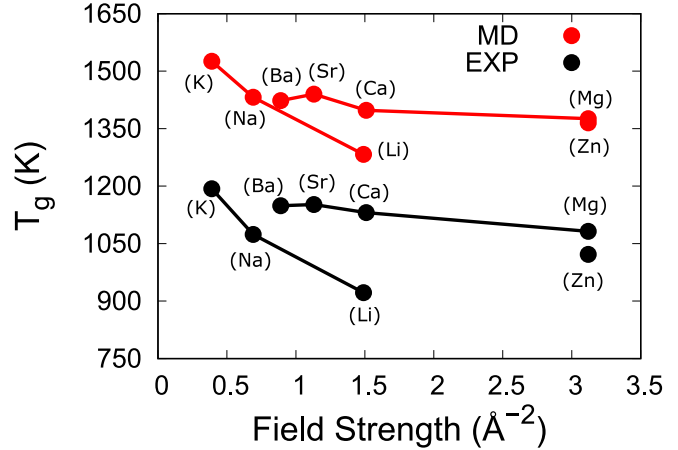}
\caption{Glass transition temperature as a function of cations field strength compared with experimental values found in ref \cite{Weigel2016}.}
\label{fig:TgFS}
\end{figure}

\textbf{\textit{Results.---}}The glass transition temperature herein was obtained from the slope break between high- and low-temperature variations of the total energy (E$_{t}$(T)) versus temperature. Fig. \ref{fig:TgFS} show T$_{g}$ values as a function of FS obtained in the present simulations together with the experimental values available in ref \cite{Weigel2016}. As it can be observed, there is a qualitative agreement in the behavior of T$_g$ with FS from our simulations and experimental studies. Also, we can see that the T$_g$ values obtained from the simulations are overestimated by around 300 K compared to experiments. This behavior is usually observed in MD simulations and is generally attributed to the very high cooling rates used in the glass preparation using MD \cite{Buchholz2002, Vollmayr1996, TAN2004, Atila2019a, Atila2019b}. Besides that, this could also be due to the simulation setup used to obtain bulk glass. The usage of PBC lead to an infinite system without any surfaces. Due to the absence of surfaces, the glasses obtained from computer simulations are usually undergo a supercooling, thus giving an effective temperature higher than the experimental one \cite{alvares2020}
The FS increases along the series K$^{+}$ $<$ Na$^{+}$ $<$ Li$^{+}$ and Ba$^{2+}$ $<$ Sr$^{2+}$ $<$ Ca$^{2+}$ $<$ Mg$^{2+}$ while FS of Zn$^{2+}$ is equivalent to that of $Mg^{2+}$. We also observe from Fig. \ref{fig:TgFS} that T$_g$ decreases with increasing FS for both alkali and alkaline earth system, with a more pronounced decrease for alkali aluminosilicate glasses. The variation of T$_{g}$ as a function of FS is found to follow Li $<$ Na $<$ K, and Zn $<$ Mg $<$ Ca $<$ Ba $<$ Sr.
\begin{figure}[ht!]
\centering
\includegraphics[width=\columnwidth]{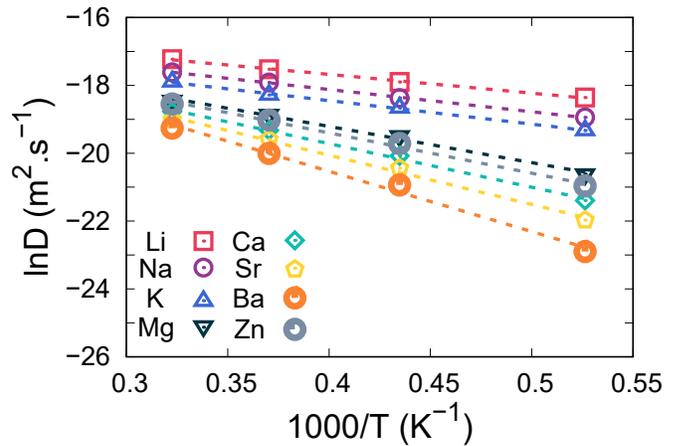}
\caption{Diffusion coefficients of Li, Na, K, Mg, Ca, Sr, Ba, and Zn as a function of temperature for the studied aluminosilicate glasses. The symbols represent the simulated data points, and the lines are fit to the Arrhenius law. (a) shows data for alkali aluminosilicate glasses.}
\label{fig:DiffTMP}
\end{figure}
Since the vitrification is a kinetic process where the dynamics of the melts play an important role, we focus in the following on the diffusion behavior at a temperature range above T$_g$ to explain the behavior of T$_g$. The mean-squared displacement (MSD) given by $MSD = \langle |r(t)- r(0)|^2 \rangle$ was used to investigate the dynamical properties of the studied system \cite{Bauchy2013b}. The MSD was calculated using trajectories under NVT runs for 60 ns and using a time step of 2 fs at each temperature (1900 - 3100 K) (see supplementary materials Fig. S1-S4 \cite{SuppMat}). Fig. S1-S4 shows that MSD values of network formers are negligible compared to values of alkali ions, we can say that alkali ions migrate rapidly in a quasi-frozen network where Si, Al and O ions will move slightly or shuffle (compared to alkali ions) to accommodate this fast migration. While alkaline earth ions move in a dynamic network since MSD of network formers (Al, Si and O) are comparable to MSD of modifiers. The diffusion coefficient D was obtained using Einstein's equation $D = \lim_{t\to\infty} \langle |r(t)- r(0)|^2 \rangle/6t$, and averaged over the last 200 ps of each run. The length of the simulation time used in this study is long enough to get a good estimate of the diffusion coefficient.
\begin{figure*}[ht!]
\centering
\includegraphics[width=0.9\textwidth]{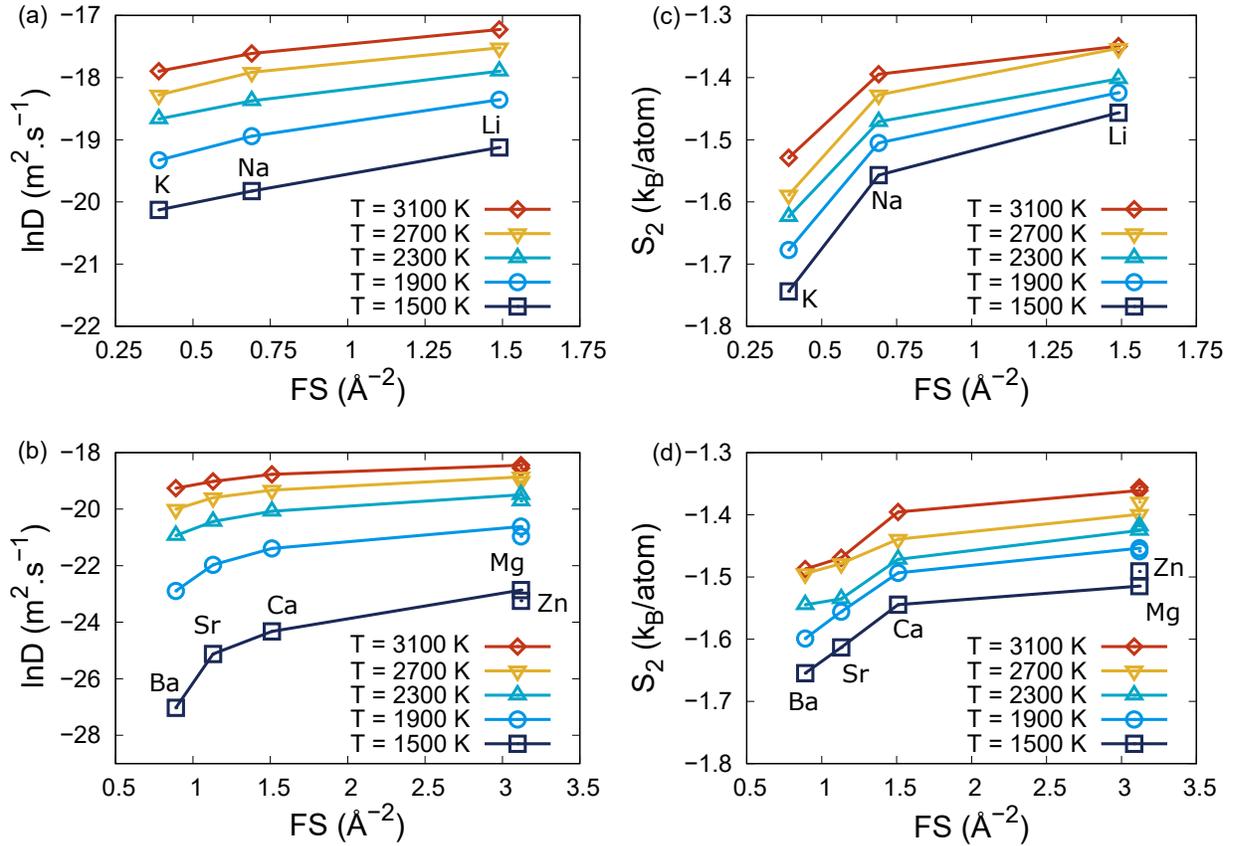}
\caption{Diffusion coefficients of Li, Na, K, Mg, Ca, Sr, Ba, and Zn as a function of charge balancing cations field strength for the studied aluminosilicate glasses for different temperatures. (a) shows data for alkali aluminosilicate glasses, and (b) shows data for alkaline earth and Zn aluminosilicate glasses. (c) and (d) show the two-body excess entropy (S$_2$/N) in the system normalized by the total number of atoms (N) in the system for alkali and alkaline earth aluminosilicate glasses respectively. The symbols represent the simulated data points, and the lines are guide to the eye.}
\label{fig:Diff_FS}
\end{figure*}
Figure \ref{fig:DiffTMP} shows the diffusion coefficients D$_X$ (where X stands for charge balancing cations) as a function of temperature in the range between 1900 - 3100 K. The partial diffusion coefficients of Si, Al, and O atoms are shown in supplementary materials Fig. S5 \cite{SuppMat}.\\
We see clearly from Fig. \ref{fig:DiffTMP} that diffusion data follows the Arrhenius behavior $\textit{D}_X = D_0exp(-\Delta E_a/k_BT)$ where D$_X$ is the diffusion coefficient of the charge balancing cations, D$_0$ is the pre-exponential factor, $\Delta$E$_a$ the activation energy barrier for self-diffusion, k$_B$ denotes Boltzmann constant, and T the temperature. The calculated diffusion coefficient from our simulations are in a good agreement with previously reported MD simulations \cite{Zhao2020, Bauchy2017, Bauchy2013}, and experimental results \cite{Novikov2017, Wu2011}. Also, knowing that the dynamical properties of liquids are very sensitive to the interaction potential \cite{Bauchy2011a}, the agreement of the data from our simulations with other MD simulations and available experimental data reinforces the reliability of our data. The alkali elements show a higher diffusion as compared to alkaline earth elements in the studied aluminosilicate melts. This indicates that the elements with a higher diffusion coefficient can diffuse longer distances and explore larger regions in the configurational space.

Fig. \ref{fig:Diff_FS} (a) and (b) show the variation of the diffusion coefficient as a function of the charge balancing cations field strength at different temperatures. As can be noticed, the self-diffusion coefficient scales with the charge balancing field strength. This behavior could be due to several factors, and to further analyze this behavior, we refer to the entropy. In general, accurate computation of the entropy is extremely costly. Thus, an expression that gives approximately the entropy should be enough to show how different cations can affect the structure of the aluminosilicate melts and glasses. This expression was derived from the expansion of the configurational entropy in terms of multibody correlation functions given by Kirkwood's factorization S = S$_{id}$ + S$_{2}$ + S$_{3}$ + ..., where S$_{id}$ is the ideal gas contribution, S$_2$ is the pair excess entropy, S$_3$ is the three-body excess entropy, and other higher terms \cite{Tanaka2019, Nettleton1958, Giuffr2010, Piaggi2017}. We use here the term S$_2$ to take only the two-body (pair) excess entropy, which is given by
\begin{equation}
\label{eq:entropy}
S_2 = -2\pi \rho k_b \int_0 ^{\infty}\left[g(r)ln g(r) - g(r) + 1\right]r^2dr.
\end{equation}
$\rho$ is the density of the system, g(r) is the radial distribution function, and k$_b$ is the Boltzmann constant.

The two-body excess entropy is plotted in Fig. \ref{fig:Diff_FS} (c) and (d) for both alkali and alkaline earth aluminosilicate systems respectively at a temperature range between 1500 - 3100 K. It is worth stressing that higher values of S$_2$ indicates higher disorder in the system. As depicted in Fig. \ref{fig:Diff_FS} (c and d) S$_2$ correlate positively with the charge balancing cations field strength and gives an indication that the systems with high FS cations are more disordered than those with low FS cations.

Additionally, by fitting lnD over 1000/T data to the Arrhenius equation, we can obtain values of the diffusion activation energy barriers $\Delta$ E$_a$ and the pre-exponential factor D$_0$. Although to obtain an accurate values of the activation energies and D$_0$, the temperature range for the fitting should be carefully chosen. The activation energies were obtained by fitting the simulation data to the Arrhenius equation in the temperature range between 1900 and 3100 K.

Figure \ref{fig:D0_Ea_CN_FS} (a) shows the activation energies for ions self-diffusion as a function of FS (see supplementary materials Fig. S6, Table. S1, and Table. S2 \cite{SuppMat} for numerical data of E$_a$ and lnD$_0$ of all elements). The activation energies for the diffusion of O, Si, and Al atoms are shown in supplementary materials Fig. S2.\\ 
As we can see, the activation energy decreases with FS for all systems, as expected from the increase of the diffusion coefficient with FS (see Fig. \ref{fig:Diff_FS}). In the same figure, the pre-exponential factor D$_0$ is also plotted as a function of FS. For the aluminosilicate systems containing alkaline earth or Zn, we noticed a striking similarity in the behavior of D$_0$ and E$_a$ as a function of FS, while for the aluminosilicate glasses containing alkali cations, D$_0$ increases with increasing FS.

\textbf{\textit{Discussion.---}}For the present glass systems, T$_g$ depends on the FS of the non-network former cations; thus, it depends directly on the radius of the charge balancing cations. The same trends have been reported by Romano \textit{et al.} \cite{ROMANO2001} in the behavior of the viscosity of XAlSi$_3$O$_8$ melts in the temperature region around T$_{g}$, where X stands for (Li, Na, K, Ca$_{0.5}$, Mg$_{0.5}$). It is also known that in aluminosilicate, T---O (T = Si or Al) bonds are the strongest. The introduction of a modifier cation induces a perturbation to these bonds in the form of competition between Si, Al, and X atoms to form bonds with oxygen atoms (X is the non-framework cation). Increasing the field strength of modifying cations leads to an increase of the perturbation in the glass network and increases the probability of the non-framework cations to form bonds with oxygen atoms \cite{Atila2019b, Navrotsky1985}. Furthermore, low FS charge balancing cations enhance the thermal stability of aluminosilicate melt structures. They do not strongly polarize the bridging oxygen and only weakly perturb the aluminosilicate structure. On the contrary, high FS cations form relatively strong bonds with the bridging oxygen that perturb the aluminosilicate structure by narrowing the T---O---T bond angle and lengthening the T---O bonds \cite{Atila2019b, hess1991role}.

The higher mobility of the charge balancing cations could be one of the reasons to explain the decrease in the glass transition temperature for the charge balanced aluminosilicate glasses. The diffusion coefficient scales with field strength as follows; Li $>$ Na $>$ K $>$ Mg $>$ Zn $>$ Ca $>$ Sr $>$ Ba, indicating that atoms which have a higher field strength (e.g., Li, Mg and/ or Zn) diffuse faster than those with lower field strength. The diffusion coefficients of the alkaline-earth aluminosilicate liquids are lower than the corresponding diffusion coefficient of the alkali aluminosilicate liquids. This can be due to the fragile nature of alkaline-earth aluminosilicate liquids (see supplementary materials Fig. S8 \cite{SuppMat}). For alkali aluminosilicate melts, the activation energy for self-diffusion decreases with increasing FS while the pre-exponential factor (D$_0$) shows an increase indicating higher jump frequency of the alkali cation. This is because increasing FS is manifested by a decrease of the cation size leading to lower mobility of the modifier cation. From another side, we noticed that when the FS increases, the coordination number decreases for both alkali and alkaline-earth cations, as shown in Fig. \ref{fig:D0_Ea_CN_FS} (b) allowing more freedom for low-coordinated cations, which is corroborated by the increase in the diffusion coefficient.
\begin{figure*}[ht!]
\centering
\includegraphics[width=0.95\textwidth]{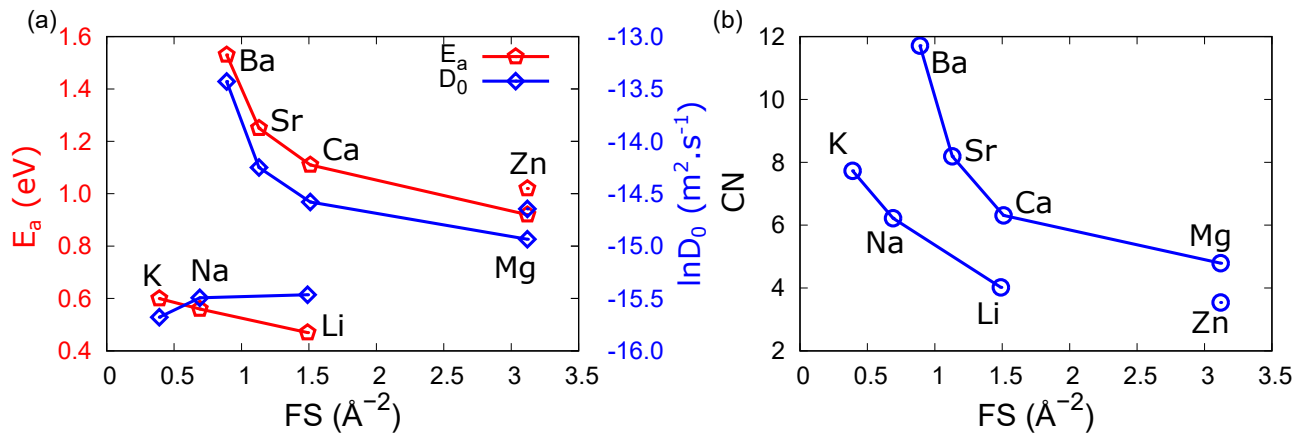}
\caption{(a) Activation energies E$_a$ and per-exponential factor D$_0$, of Li, Na, K, Mg, Ca, Sr, Ba, and Zn as a function of charge balancing cations field strength for the studied aluminosilicates. (b) mean coordination number of the charge balancing cations to oxygen as a function of FS in the studied aluminosilicates at 300 K. The symbols represent the simulated data points, and the lines are guide to the eye.}
\label{fig:D0_Ea_CN_FS}
\end{figure*}
As discussed above, Alkali cations have higher diffusion coefficients and smaller activation energy barriers, so there are faster and easier to diffuse. The pre-exponential factor D$_0$ for these elements increases with increasing FS meaning that high FS cations, can explore more configurational space and relax toward lower energy minima, thus a lower T$_g$. As for alkaline earth cations, both the activation energies for self-diffusion and D$_0$ decreases with increasing FS, indicating that even if the activation energy barriers for self-diffusion are smaller for high FS elements, the jump frequency is lower which lead to slow relaxation of the systems. However, the values of T$_g$ of potassium aluminosilicate glass are higher than T$_g$ of the magnesium aluminosilicate glass. This could be attributed partially to the high coordination state of potassium as compared to magnesium and partially to the cation molar mass (potassium is heavier than magnesium). This fact could also explain the higher T$_g$ values of alkaline earth AS glasses. The low energy barriers for self-diffusion found in those melts give to the systems the freedom to relax more towards lower energy minima resulting to lower T$_g$ values as compared to other systems with lower field strength and higher energy barriers for self-diffusion. This explanation is supported by the two-body excess entropy behavior as it is known in literature that disorder enhances the ability of atoms to diffuse \cite{ren2016structural, Avramov1988, kirchheim1983, Swenson1996}.

The fragility is a scalar to quantifies the rate at which any dynamical quantity such as viscosity or diffusion grows with temperature \cite{Angell1998, Ito1999, Banerjee2016}. The fragility of the liquid was computed from the Arrhenius behavior as in Refs. \cite{Bauchy2013b, Micoulaut2010} (see supplementary materials Fig. S9\cite{SuppMat}). The values calculated here are in good agreement with the values reported in many experiments \cite{Moesgaard2009, Bechgaard2017, ROMANO2001}. The high fragile nature of the alkaline earth aluminosilicate melts may result from the differences in the diffusion in these systems as compared to the alkali aluminosilicate melts. In contrast to the glass-forming melts with high fragility, the low fragility glass-forming melts show a relatively broad glass transition temperature range as they exhibit relatively a slow change in the viscosity with temperature which give more time for the glass forming melts to relax its structure.

As we have noticed from the results presented herein, the pre-exponential factor D$_0$ and diffusion energy barrier E$_A$ have an important contribution to the diffusion coefficient as for the activation energy governed by the ionic size and masse of the cation and by the structural configuration of the host medium. From the fitting of the diffusion data to the Arrhenius equation, we determine the values of D$_0$ and E$_A$ and its contributions to the diffusion coefficients of the studied glasses. We found that the pre-exponential factor D$_0$ plays a significant role along with the activation energy barrier E$_a$ in determining the magnitude of D, and it shows a correlation with FS as for the activation energies. Self-diffusion coefficient D can be expressed as
\begin{center}
\begin{equation}
\label{eq:Diffentropy}
\begin{split}
D = D^*exp(-\dfrac{G}{k_BT}) = D^*exp(\dfrac{S}{k_B})exp(-\dfrac{H}{kT})\\
= D_0exp(-\dfrac{H}{k_BT})
\end{split} 
\end{equation}
\end{center}
where $D_0 = D^*exp(\dfrac{S}{k_B})$, G, H, and S denote the Gibbs free energy, enthalpy, and entropy of activation of the self-diffusion process, respectively, the pre-exponential factor and activation energy are dominated by entropy and enthalpy contributions. In the charge-balanced aluminosilicate glasses, the self-diffusion of cations is dominated by the base glass, which has different glass structures, as shown by simulation results. This difference in the structure leads to different configurational and vibrational entropy and hence different activation entropy leading to different pre-exponential factors of diffusion. The pair excess entropy shows qualitatively that the configurational entropy of the studied melts is affected remarkably by the charge balancing cations.

\textbf{\textit{Conclusion.---}}Molecular dynamics simulations were performed to understand the origins of the anomalous behavior of T$_g$ as a function of FS and the effect of the charge balancing cations on the diffusion behavior in the aluminosilicates. The results showed that the  origins of this anomaly are linked to the diffusion of the charge balancing cations. The high FS cations diffuse faster than the low FS cations, and this behavior was attributed to the high pair excess entropy, which also indicates a high disorder. This high diffusion of the charge balancing cations allows for a relaxation of the structure, thus, exploring low energy minima, which leads to a low T$_g$. Additionally, the melt fragility showed that AS glasses with high field strength tend to be strong glass forming melts (low fragility index) compared to the ones with high FS; the fragility of the alkaline earth aluminosilicate melts is also responsible for the differences in the diffusion in these systems as compared to the alkali aluminosilicate melts. Altogether the results presented in this letter will deepen our understanding of the role of different modifiers on the diffusion behavior in silicate and aluminosilicate (as the insight given here is expected to hold also for binary silicate systems) and help in getting an atomic-scale understanding of the glass transition.

\bibliographystyle{apsrev4-2}
\bibliography{main}
\end{document}